# Optically-Nonactive Assorted Helices Array with Interchangeable Magnetic/Electric Resonance

Xiang Xiong[1], Xiao-Chun Chen[1], Mu Wang[1,*], Ru-Wen Peng[1], Da-Jun Shu[1] and Cheng Sun[2]

[1]*National Laboratory of Solid State Microstructures and Department of Physics, Nanjing University, Nanjing 210093, China*

[2]*Department of Mechanical Engineering, Northwestern University, IL 60208-3111, USA*

We report here the designing of optically-nonactive metamaterial by assembling metallic helices with different chirality. With linearly polarized incident light, pure electric or magnetic resonance can be selectively realized, which leads to negative permittivity or negative permeability accordingly. Further, we show that pure electric or magnetic resonance can be interchanged at the same frequency band by merely changing the polarization of incident light for 90 degrees. This design demonstrates a unique approach to construct metamaterial.

[*]To whom correspondence should be addressed. E-mail: muwang@nju.edu.cn

The interest to construct metamaterial has being promoted by its unique optical properties, such as negative refractive index,[1-4] ultrahigh spatial resolution,[5-6] invisibility cloaking,[7-8] and optical magnetics.[9-10] These fantastic properties facilitate potential applications in extraordinary optical transmission,[11-15] microscopy,[16] and antennas,[17] etc.. Among the massive researches in this rapidly developing area, one of the intensively studied subjects is chiral metamaterial, which offers an alternative approach to realize negative refractive index.[18-23] Conventional designing of the material with negative refractive index follows the idea to combine a resonant magnetic structure with metal that provides a "background" of negative permittivity in a broad spectral range, including the wavelength where magnetic response associated with the specific structure occur.[1-2] With different approach, chirality of the system helps to suppress the refractive index of light with one handedness, and increases the refractive index of light with the other handedness.[18-23] With sufficiently strong chirality, a negative refractive index is realized. So far, various structures, such as cross-wires,[20] twisted rosettes,[21] interlocked split-ring resonators[23] and U-shaped resonators,[22] have being proposed to construct the negative-refractive-indexed chiral metamaterials.

In chiral metamaterial the strongly coupled electric and magnetic dipoles are simultaneously excited. In previous structures, there usually exists an angle between electric and magnetic dipoles, which prevents the fully use of the induced dipoles. We once proposed an assembly of double-layered metallic U-shaped resonators with two resonant frequencies $\omega_H$ and $\omega_L$.[22, 24] The effective induced electric and magnetic dipoles, which originate from the specific distribution of induced surface electric current upon the illumination of incident light, are collinear at the same frequency. Consequently, for left circularly polarized incident light, negative refractive index occurs at $\omega_H$, whereas for right circularly polarized incident light it occurs at $\omega_L$. Despite of the efficient use of the electric and magnetic dipoles in that approach, however, the coupling of the building blocks between the different layers is very strong, which is an unfavorable factor in constructing three-dimensional bulk metamaterial. One possible approach to solve this problem is to introduce helix structure.[25] The advantage of helix structure is that collinear electric and magnetic dipoles in parallel or anti-parallel directions can be excited in both left-handed and right-handed helices. In addition, by combining helices with different chirality, it is possible to construct an optically nonactive metamaterial with assorted helices array.

Here we show that with linearly polarized incident light, collinear parallel and antiparallel electric and magnetic dipoles are induced in the helix array, and pure electric or magnetic resonance can be selectively realized with specific combination of helices with different chirality. Accordingly, negative permittivity or negative permeability can be achieved. We demonstrate further that the pure electric or magnetic resonance can be interchanged at the same frequency band with this assembly by merely changing the polarization of incident light for 90 degrees.

The elementary building block is a uniaxial gold helix $H_1$ with three turns, as illustrated in Fig. 1(a). The axis of the helix is along $x$-axis and the two endpoints locate in $x$-$y$ plane. The incident wave vector is set along $z$-direction and the polarization of incident light can be switched either in $x$ or in $y$ direction. In the simulation, the permittivity of gold in the infrared regime is based on the Drude model, $\varepsilon(\omega)=1-\omega_p^2/(\omega^2+i\omega_\tau\omega)$, where $\omega_p$ is the plasma frequency, and $\omega_\tau$ is the damping constant. For gold, these characteristic frequencies are taken as $\omega_p$=1.37× $10^4$ THz, and $\omega_\tau$ =40.84 THz.[26] Figure 1(b) shows the transmission coefficients of the array of helix unit shown in Fig. 1(a). The distance between two neighboring helices in *x* and *y* directions is 0.25 $\mu$m. The resonant dips in the transmission $t_\parallel$, where the polarization of input and output waves is parallel, can be detected at 1200 cm$^{-1}$ for both *x*-polarized ($t_{x\parallel}$) and *y*-polarized ($t_{y\parallel}$) incidence. It should be noted that the helix possesses intrinsic chirality and the helix array shows optical activity. The chiral behavior of the helix array rotates the polarization of incident light and converts a portion of energy from one polarization to the other. Hence resonance peaks of perpendicular polarization of input and output light ($t_\perp$) can be detected in the transmission around 1200 cm$^{-1}$ for both *x* polarization ($t_{x\perp}$) and *y* polarization ($t_{y\perp}$). Simulation results show that $t_{x\perp}$ and $t_{y\perp}$ are almost identical in Fig. 1(b). It should be noted that there exist higher order of resonances in the helix structure, where the induced surface current flows with a complicated pattern (i.e., the current flows in different directions on different section of the helix). Here we focus on the lowest frequency resonance of the structure (1200 cm$^{-1}$), where the induced surface current flows in the same direction, as shown in Fig. 1(c) and (d). The resonance mode of the surface current in the helix is polarization independent. The surface current shown in Fig. 1(c) essentially flows along -*x*-direction, which corresponds to an effective electric field *E'* in –*x* direction [as indicated by green arrow in Fig. 1(d)]. In other word, an effective electric dipole along –*x* direction is induced. On the other hand, the surface current along the helix H$_1$ forms a loop structure. The curl integration along the projection in *y*-*z* plane is nonzero, which establishes an induced magnetic field *H'* in –*x* direction. This indicates that an effective magnetic dipole along –*x* direction is induced at the same time [as indicated by blue arrow in Fig. 1 (d)].

In Fig. 1(d) the induced electric field and magnetic field are both along $-x$ direction, i.e., the induced electric and magnetic dipoles point to the same direction, which is the most favorable in generating optical activity. For a helix with the opposite chirality, $H_2$ (see in Fig.1(e)), collinear electric dipoles and magnetic dipoles can also be excited with both *x*-polarized and *y*-polarized incidence, yet the induced electric and magnetic dipoles point to the opposite directions, as shown in Fig. 1(f).

It should be pointed out that the induced surface electric current through the helix contributes not only to the electric/magnetic field along the helix axis, but also to some small components along the other directions while the current is rotating in the helix. Those undesired components of electric/magnetic field can be eliminated by superposing helix with different handedness.[27] Our approach is illustrated in Fig. 2. The helix $H_1$ locates in the second quadrant (-*x*,+*y*), with helix axis along *x*-axis and the two endpoints locating in *x-y* plane. By taking *y-z* plane as the mirror plane, a mirror image of $H_1$, which has different chirality as $H_1$, is generated and denoted as $H_2$. Thereafter, by taking *x-z* plane as the mirror plane, two other mirror images, $H_2$' and $H_1$', which are the mirror images of $H_1$ and $H_2$, respectively, are introduced. These four helices are so arranged that *x* and *y* directions are the symmetry axes of the unit cell. This unit cell is periodically reproduced, and the lattice parameter is 0.25 $\mu$m in both *x* and *y* directions. When the incident light propagates along *z* direction, the induced resonant surface current at 1200 cm$^{-1}$ is illustrated in Fig. 2(a) (*x*-polarized incidence) and in Fig. 2(b) (*y*-polarized incidence), respectively. With *x*-polarized incident light, the induced surface current in both $H_1$ ($H_1$') and $H_2$ ($H_2$') moves towards *x* direction in general (Fig. 2(a)), and the induced electric dipoles in $H_1$ ($H_1$') and $H_2$ ($H_2$') are both in *x* direction (Fig. 2(c)). The induced magnetic dipoles of $H_1$ ($H_1$') are along *x* direction and that of $H_2$ ($H_2$') are along $-x$ direction. In other words, the electric dipoles of $H_1$ ($H_1$') and $H_2$ ($H_2$') are parallel while the magnetic dipoles are antiparallel. Therefore, with such a combination of helices, as illustrated in Fig. 2(c), a pure electric resonance is achieved when the incident light is *x*-polarized.

For *y*-polarized incidence, as illustrated in Fig. 2(d), the induced surface currents in $H_1$ ($H_1$') and $H_2$ ($H_2$') move antiparallel along *x* direction. So the induced electric dipoles are in opposite directions, whereas the induced magnetic dipoles are both along *x* direction. So for *y*-polarized incident light, with such a combination of helices, a pure magnetic resonance is established.

Due to the high symmetry of the unit cell, when the incident light is *x*- or *y*-polarized, both the tensors of permeability and permittivity become diagonal. Therefore, a metamaterial constructed with the combination of helices $H_1$ ($H_1$') and $H_2$ ($H_2$') will be optically nonactive.

Transmission and reflection coefficients for both *x*- and *y*-polarized incidence are calculated respectively in Fig. 3(a) and (b). Resonant dips in transmissions and peaks in reflections appear in the region 1100 cm$^{-1}$-1300 cm$^{-1}$. Simulation confirms that when the incident light is *x*- and *y*- polarized, no energy can be detected in the perpendicular polarization direction. Our calculation also reveals that the transmission and reflection along *z* and –*z* directions are identical. The retrieval method based on S-parameters[28] has been applied to achieve effective permittivity and permeability, in which the complex transmission and reflection coefficients describe the scattered wave. Figure 3(c) and 3(d) illustrate the retrieved effective permittivity ($\varepsilon_{eff}$) and permeability ($\mu_{eff}$) of the structure. Pure electric resonance (characterized by an evident jump in $\varepsilon_{eff}$) occurs in 1100 cm$^{-1}$-1300 cm$^{-1}$ for *x*-polarized incidence, and pure magnetic resonance (characterized by an evident jump in $\mu_{eff}$) occurs for *y*-polarized incidence at the same frequency. This means that by changing the polarization of incident light, the electric and magnetic resonances of the system can be interchanged in the same frequency band.

In order to confirm that pure electric and magnetic resonances really occur, we calculate the distribution of electric and magnetic fields along *x*-direction at the resonant frequencies with different incident polarization, as illustrated in Fig. 4. The *x-y* plane is used as the cutting plane. The dash lines divide the unit cell into four regions, in each region a helix with different chirality is placed (the arrangement corresponds exactly to that shown Fig. 2(a)). For *x*-polarized incidence (the scenario of Fig. 2(a)), the electric field in all the four parts of the unit cell is in phase (Fig. 4(a)), whereas the magnetic field is in antiphase and is consequently canceled (Fig. 4(b)). For *y*-polarized incidence (the scenario of Fig. 2(b)), the electric field in the four parts of the unit cell is canceled (Fig. 4(c)), whereas the magnetic filed in these parts is in phase and is hence summed up (Fig. 4(d)). Therefore the field distribution at resonance shown in Fig. 4 indicates that pure electric resonance occurs when the incident light is *x*-polarized, while pure magnetic resonance occurs to the same structure when the incident light is *y*-polarized.

It should be noted that the frequency at which the electric or magnetic resonance occurs depends on the electromagnetic interactions between the units. One may easily find out that by resuming the other parameters, the resonance of helix array shifts to the lower frequency when the length of the helix is elongated. The resonance frequency shifts to higher frequency when the helix is shortened. The resonance can also be tuned by changing the spatial periodicity of the array and the geometrical parameters of helix. It is expected to be a universal feature that pure electric and magnetic resonances, and hence negative permittivity and negative permeability, can be switched at the same frequency by changing the polarization of incident light for $90^{o}$. Our design provides an interest approach to realize simultaneous magnetic and electric resonances and constructing optically nonactive metamaterial with chiral building blocks.

This work was supported by grants from the National Natural Science Foundation of China (Grant Nos. 50972057, 10874068, 11034005, and 61077023), the State Key Program for Basic Research from the Ministry of Science and Technology of China (Grant Nos. 2010CB630705), and partly by Jiangsu Province (Grant No. BK2008012).

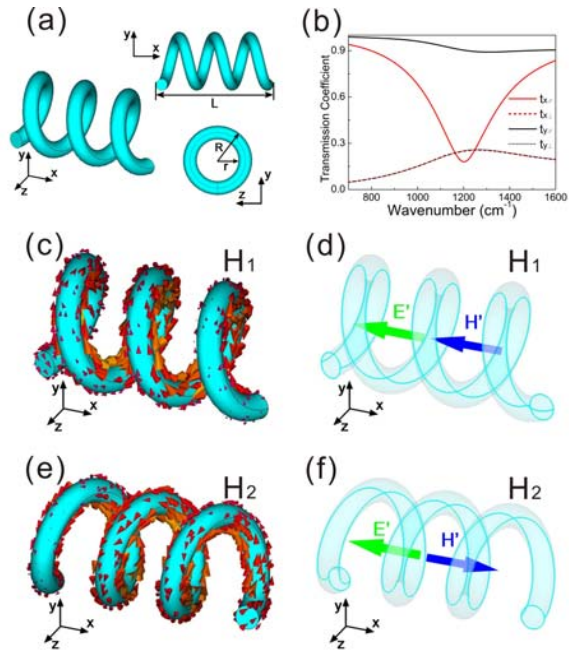

FIG. 1. (Color online) (a) The geometrical parameters of building block $H_1$: L=1.0$\mu$m, R=0.25$\mu$m, r=0.15$\mu$m. (b) Transmission coefficients for ***x*** polarization and ***y*** polarization incident lights of $H_1$ array. (c)/(e) The calculated induced surface electric current density on $H_1$/$H_2$. (d)/(f) Schematics to show the equivalent electric and magnetic dipoles induced on $H_1$/$H_2$.

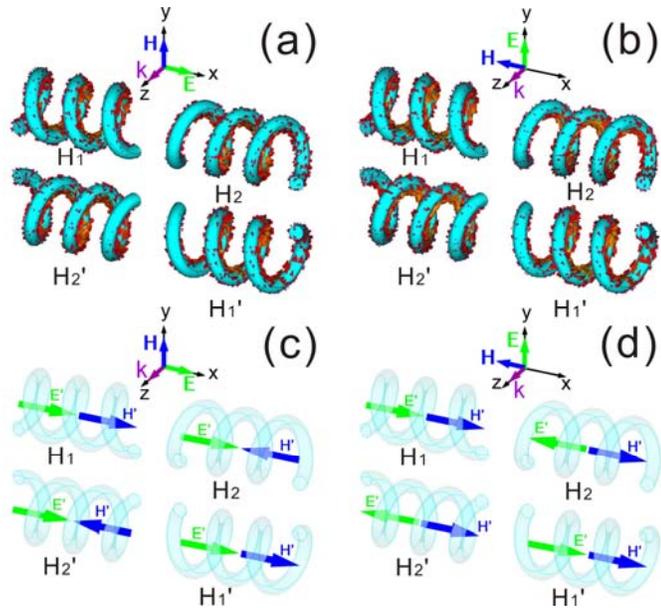

FIG. 2. (Color online) (a)-(b) The calculated induced surface electric current density on gold helix metamaterial when the incident light is polarized in $x$ (a) and $y$ (b) directions, respectively. (c)-(d) The schematics to show the equivalent electric and magnetic dipoles induced on each building blocks when the incident light is polarized along $x$ (c) and $y$ (d) directions, respectively.

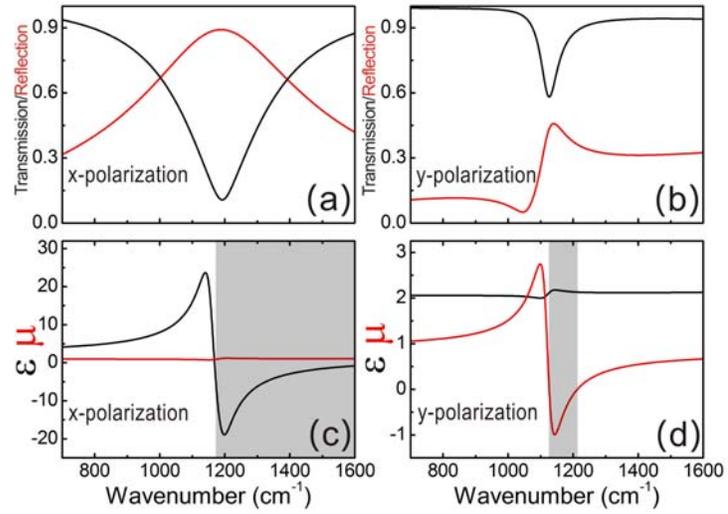

FIG. 3. (Color online) (a)-(b) The calculated transmission (black line) and reflection (red/gray line) coefficients for *x/y*-polarized incident light. (c)-(d) The retrieved effective permittivity and permeability for *x/y*-polarized incident light, respectively.

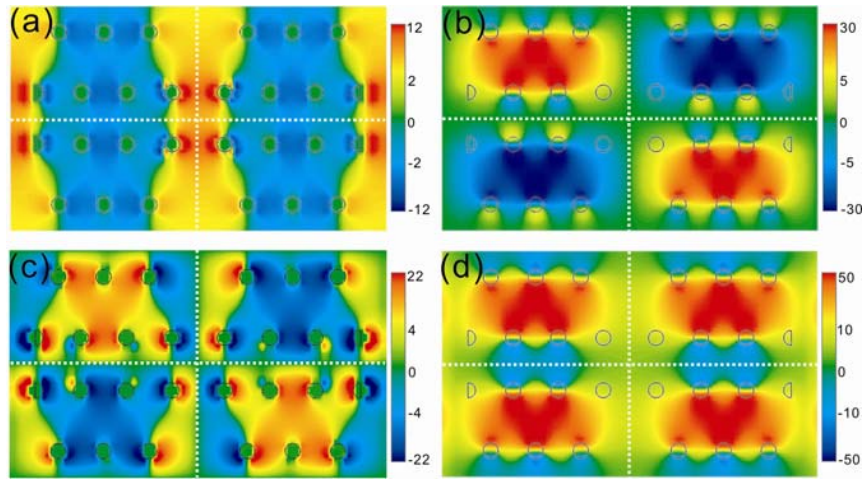

FIG. 4. (Color online) (a)-(b) The calculated distribution of electric (a) and magnetic (b) fields in $x$ direction for $x$-polarized incident light. (c)-(d) The calculated distribution of electric (c) and magnetic (d) fields in $x$ direction for $y$-polarized incident light.